\documentclass[journal=jacsat,manuscript=article]{achemso}

\usepackage{chemformula} 
\usepackage[T1]{fontenc} 

\usepackage{booktabs, tabularx, float}
\usepackage{ragged2e}

\author{Chenghao Wan}
\author{Connor Cremers}
\author{Ariana B. Höfelmann}

\affiliation[EE department]
{Department of Electrical Engineering, Stanford University, Stanford, CA 94305, USA}
\author{Zhennan Ru}

\affiliation[MSE department]
{Department of Materials Science \& Engineering, Stanford University, Stanford, CA 94305, USA}

\author{Calvin H. Lin}
\affiliation[EE department]
{Department of Electrical Engineering, Stanford University, Stanford, CA 94305, USA}

\author{Kesha N. Tamakuwala}
\affiliation[Chemistry department]
{Department of Chemistry, Stanford University, Stanford, CA 94305, USA}

\author{Dolly Mantle}
\affiliation[Mechanical department]
{Department of Mechanical Engineering, Stanford University, Stanford, CA 94305, USA}

\author{Pinak Mohapatra}
\affiliation[EE department]
{Department of Electrical Engineering, Stanford University, Stanford, CA 94305, USA}

\author{Juan Rivas-Davila}
\affiliation[EE department]
{Department of Electrical Engineering, Stanford University, Stanford, CA 94305, USA}

\author{Matthew W. Kanan}
\affiliation[Chemistry department]
{Department of Chemistry, Stanford University, Stanford, CA 94305, USA}

\author{Jonathan A. Fan}
\affiliation[EE department]
{Department of Electrical Engineering, Stanford University, Stanford, CA 94305, USA}

\email{jonfan@stanford.edu}
\phone{}
\fax{}

\title
  {Scale up analysis of inductively heated metamaterial reactors}

\abbreviations{IR,NMR,UV}

\begin{document}







\begin{abstract}
 Inductively heated metamaterial reactors, which utilize an open cell lattice baffle structure as a heating susceptor for magnetic induction, are promising candidates for scaled electrified thermochemical reactor operation due to their ability to support volumetric heating profiles and enhanced heat transfer properties. In this work, we present a systematic scale up analysis of inductive metamaterial reactors where we utilize a combination of analytic modeling, numerical simulations, and experiments to project the capabilities and performance of scaled reactors. We use reverse water gas shift as a model reaction system and show that for reactor configurations featuring a uniform metamaterial susceptor, the total system efficiency increases with scale. However, the throughput of these scaled reactors is limited by radial temperature gradients. We further show this bottleneck can be overcome by tailoring the radial effective conductivity profile of the susceptor, which can enable scaled reactors with nearly ideal plug flow-like capabilities. These concepts provide a pathway towards scaled electrified thermochemical reactors with optimal chemical conversion capabilities.
\end{abstract}

\section{Introduction}

The electrification of thermochemical reactors serves as a promising route to decarbonizing chemical manufacturing processes and enhancing the capabilities of chemical reactors.\cite{Thiel2021, VanGeem2019} To date, proof-of-concept demonstrations of various electrified reactor heating methods have been explored, including those based on resistive joule heating, \cite{Badakhsh2021, Dong2023, Wismann2019, From2024} microwaves,\cite{Hou2021, Kappe2013, BakerFales2023, Sundaramoorthy2025} thermal plasmas,\cite{Morais2023, Akande2022, Akay2020, Mohamed2025} and magnetic induction.\cite{Ceylan2008, Scarfiello2021, Faure2021, Teel2025, Yan2024}  These studies have shown that electrified heating methods can support enhanced heat transfer rates to catalytic sites beyond the capabilities of fossil fuel-based heating methods, enabling the process intensification of highly endothermic reactions.\cite{Kim2023, Wang2024, Lucia2014}   Electrified heating methods have also been shown to support the selective heating of reactants and catalysts, enhancing the selectivity and conversion of reactants to products.\cite{Zhao2017, Deng2025}  

While electrified thermochemical reactors hold great promise as  translational clean energy technologies, scalability remains a key challenge.\cite{Mallapragada2023, Noble2024}  Resistive heating elements are highly efficient but typically have a form factor of one-dimensional tubes or wires that are challenging to adapt to three-dimensional volumes. Additionally, ensuring robust electrical contacts at high temperatures is non-trivial and often requires specialized materials. Microwave heating sources enable wireless and selective heating but are limited by the finite penetration depth of microwaves in typical reactor environments and inefficient power electronics. Thermal plasma heating can achieve extreme temperatures suitable for specialized reactions such as pyrolysis, but it is difficult to adapt to mainstream catalytic systems. With these considerations in mind, magnetic induction heating has been identified as a particularly attractive approach to scaled electrified reactor operation because it can produce volumetric heating profiles, operates with high input voltages and low input currents due to its intrinsic transformer properties, and has been demonstrated to effectively operate at megawatt power levels.\cite{Rudnev2017}  Modes of induction heating include the heating of magnetic materials by hysteresis, which can enable the selective heating of magnetic catalysts,\cite{Ceylan2011} and the more general heating of electrically conductive materials by eddy currents,\cite{Wang2019} which will be the focus here. 

Among the most advanced eddy current-based inductive reactor concepts are metamaterial reactors, which we recently proposed and where an internal structured reactor baffle susceptor is tailored with the magnetic induction frequency to enable volumetric heating and high coupling efficiencies.\cite{Lin2024}  Coupling efficiency here corresponds to the fraction of electrical energy inputted into the magnetic coil that converts to heat in the susceptor.  An important design consideration with metamaterial reactors is the modeling of the baffle as a homogeneous medium with an effective electrical conductivity, which enables the tractable evaluation and optimization of macroscopic reactor heating properties.  These induction heating concepts complement recent developments in structured reactors, where architected baffles have been proposed to enhance the heat transfer and hydrodynamic properties within a reactor.\cite{Ambrosetti2020, Visconti2016, Eigenberger2012, Kapteijn2022}  In an initial lab scale demonstration, we showed that a volumetric reactor metamaterial baffle comprising a conductive open-cell ceramic foam lattice could be packed with fixed bed catalytic material and facilitate the reverse water gas shift (RWGS) reaction near thermal equilibrium conversion limits.   

In this study, we investigate how the capabilities and efficiencies of inductively heated metamaterial reactors scale as the reactor size increases from lab to commercial scales.  We specifically investigate reactor metrics as a function of $\beta$, which is a linear scaling factor for the diameter and length of the reactor and the coil pitch  (Fig. 1A).  We consider $\beta$ ranging from 1 to 32, which corresponds to reactor radii ($R$) spanning 0.0094 to 0.3 meters. We show that for reactors with uniform metamaterial baffles of constant effective electrical conductivity, the baffle conductivity and induction frequency can be tailored as a function of $\beta$ to enable high coupling efficiency while maintaining a volumetric parabolic heating profile.  We then project the capabilities of scaled reactors with uniform baffles, using a multiphysics model fine-tuned with experimental lab scale reactor data, and find that conversion and throughput are ultimately limited by radial temperature gradients.  We then discuss how these radial temperature gradients can be mitigated by tailoring the radial effective conductivity profile of the metamaterial susceptor, enabling scaled reactor performance with nearly ideal plug flow-type characteristics.

\section{Results and discussion}

\subsection{Coupling efficiency scaling}

We begin our analysis by briefly reviewing the heating properties and ideal operating regimes of inductive metamaterial tubular reactors with uniform cylindrical baffles.  A more detailed discussion can be found in Ref.~\citenum{Lin2024}.  The macroscopic heating profile is dictated by the susceptor skin depth, $\delta$, which represents the penetration depth of the coil-generated alternating magnetic field into the susceptor.  It is defined to be $\delta = \sqrt{1/(\pi \sigma_{\mathrm{eff}} f \mu_0)}$, where $f$ is the induction frequency and $\sigma_{\mathrm{eff}}$ is the effective electrical conductivity of the metamaterial susceptor.  When $\delta$ is larger than $\sim R/2$, the heating profile is volumetric and approximately parabolic, and when $\delta$ is smaller than $\sim R/2$, the heating profile becomes confined to the outer circumferential edges of the reactor.  

\begin{figure}

\includegraphics[width=\textwidth]{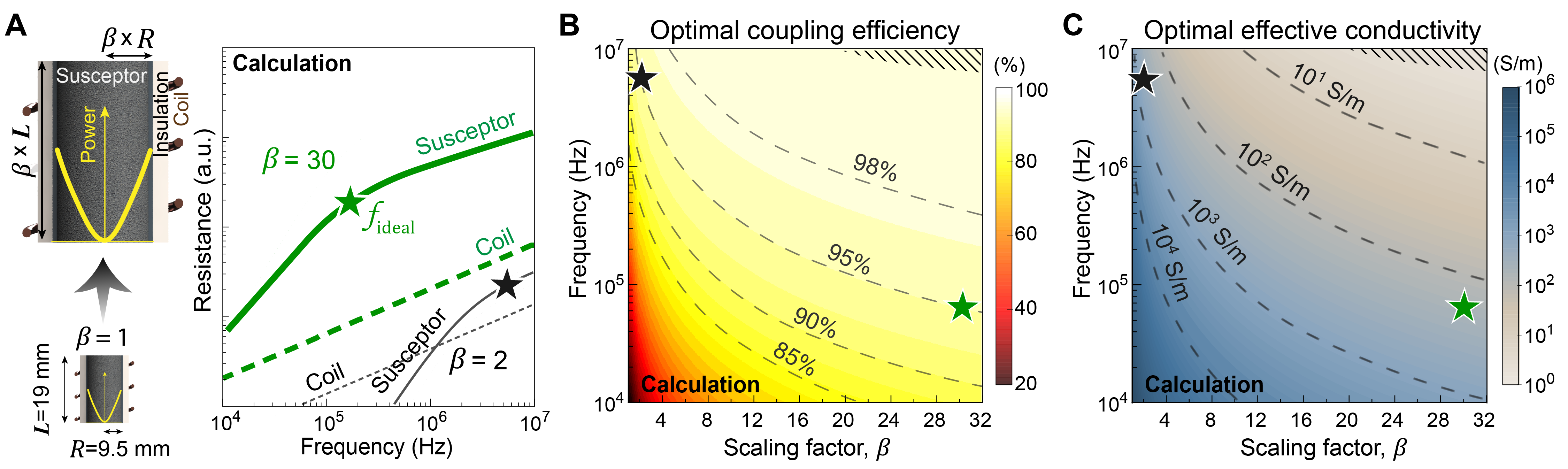}
\captionsetup{font=small}
\captionsetup{labelfont=bf}

\caption{\textbf{Coupling efficiencies of scaled metamaterial reactors with uniform $\sigma_{\mathrm{eff}}$. (A)} AC resistances of the susceptor ($R_{\mathrm{susc}}$) and coil ($R_\mathrm{coil}$) as a function of induction frequency for two different reactor sizes. The $R_{\mathrm{susc}}$ curves feature a low frequency, volumetric heating regime where $R_{\mathrm{susc}} \propto f^2$ and a high frequency, surface heating regime where $R_{\mathrm{susc}} \propto \sqrt{f}$ . The transition frequency, $f_{\mathrm{ideal}}$, separates these regimes and delineates the optimal heating condition.
\textbf{(B)} Contour plot of coupling efficiency as a function of $\beta$ and operating frequency, assuming optimal heating conditions. Regions of high coupling efficiency (over 90\%) are achieved at MHz frequencies for small $\beta$ and at kHz frequencies for large $\beta$.
\textbf{(C)} Contour plot of optimal $\sigma_{\mathrm{eff}}$ corresponding to ideal heating conditions.
}
  \label{fgr:example}
\end{figure}

The total power dissipated in the susceptor relates to the susceptor AC resistance ($R_\mathrm{susc}$), which is a function of $\delta$, $\sigma_{\mathrm{eff}}$, and $R$ and which has a frequency dependent profile shown in Fig. 1A.  The magnetic coil itself also dissipates heat and has an AC resistance, $R_\mathrm{coil}$.  More detailed expressions of these terms are in the Supplementary Sections 1 and 4.  At relatively low frequencies, $R_\mathrm{susc}$ scales as $f^2$ while $R_\mathrm{coil}$ scales as $\sqrt{f}$ and the coupling efficiency, $R_\mathrm{susc}/(R_\mathrm{coil}+R_\mathrm{susc})$, increases with frequency.\cite{Lin2024} At high frequencies, both $R_\mathrm{susc}$ and  $R_\mathrm{coil}$ scale as $\sqrt{f}$ and the coupling efficiency is capped.  The transition point between these regimes occurs at $f_{\mathrm{ideal}}$ when $\delta \sim R/2$.  This point represents the ideal induction heating condition for a particular reactor configuration, as it features a volumetric heating profile and nearly maximal coupling efficiencies.

We utilize this understanding to probe how scaled reactor systems with uniform susceptors can operate, assuming ideal heating conditions.  A contour map showing coupling efficiency for differing $\beta$ and $f$ values is shown in Fig. 1B.  The baffle $\sigma_{\mathrm{eff}}$ at each point on the map is adjusted to ensure ideal heating conditions, and the corresponding contour map of $\sigma_{\mathrm{eff}}$ is presented in Fig. 1C. Each contour plot also includes reference lines representing constant coupling efficiency (Fig. 1B) or effective conductivity (Fig. 1C) values, and the points from Fig. 1A are also plotted as specific examples.  The hatched region at the top right of the plots represents a non-operable regime where the magnetic induction coil itself is self resonant and no longer operating in the quasi-magnetostatic limit, setting frequency limits in scaled systems (See Supplementary Section 4 for more discussion).  The boundary of this region is conservatively specified, as parasitics in the coil system can practically lead to a larger non-operable regime in scaled systems.

For small reactors, coupling efficiencies greater than 90\% are possible only at megahertz frequencies.  As the reactor scales and coupling efficiency is maintained constant, $f$ follows the reference lines and reduces to values below one megahertz (Fig. 1B), during which $\sigma_{\mathrm{eff}}$ remains nearly constant (Fig. 1C). The ability for scaled reactor systems to operate with high coupling efficiencies and with $f$ in the kilohertz range is important from the standpoint of power electronics.  Megahertz frequency induction requires the use of high frequency power electronics based on wide bandgap switches and resonant power amplifier topologies, which have limited power output and are best suited for lab- and pilot-scale reactors.  At kilohertz frequencies, our scaled reactor systems can utilize silicon-based power electronics, which are a more mature technology, can support higher power levels, and can convert DC to AC power with near unity efficiency.  

The range of susceptor $\sigma_{\mathrm{eff}}$ values required for different scaled system configurations necessitates a customizable susceptor platform for tuning $\sigma_{\mathrm{eff}}$.  We show experimentally that $\sigma_{\mathrm{eff}}$ can be widely tuned by tailoring the geometry and composition of the susceptor, and that this tuning can be accurately characterized using an effective medium picture described by the Lemlich limit,\cite{Lemlich1978} which provides a simple expression relating $\sigma_{\mathrm{eff}}$ to the baffle material electrical conductivity, lattice porosity, and lattice tortuosity (Fig. 2A).  To demonstrate, we perform AC resistance measurements on open cell lattice susceptors with the same cylindrical geometry (38 mm diameter, 150 mm length) but made from different high temperature-stable materials (Haynes superalloy metal and reaction bonded silicon carbide, i.e., SiSiC), different lattice types (random foams and cubic lattices made by additive manufacturing), and porosity values ranging from 0.8 to 0.99.  The resistance curves are plotted in Fig. 2B and show that the impedance curves all have AC resistance values that initially scale as $f^2$ followed by $\sqrt{f}$, which are the same trends featured in Fig. 1A for homogeneous media.  Furthermore, the $f_{\mathrm{ideal}}$ of the curves span frequencies ranging over two orders of magnitude, and they lie on a trend line proportional to frequency, which follows the theory for homogeneous conductive cylinder induction heating.\cite{Lin2024}

\begin{figure}
\includegraphics[width=\textwidth]{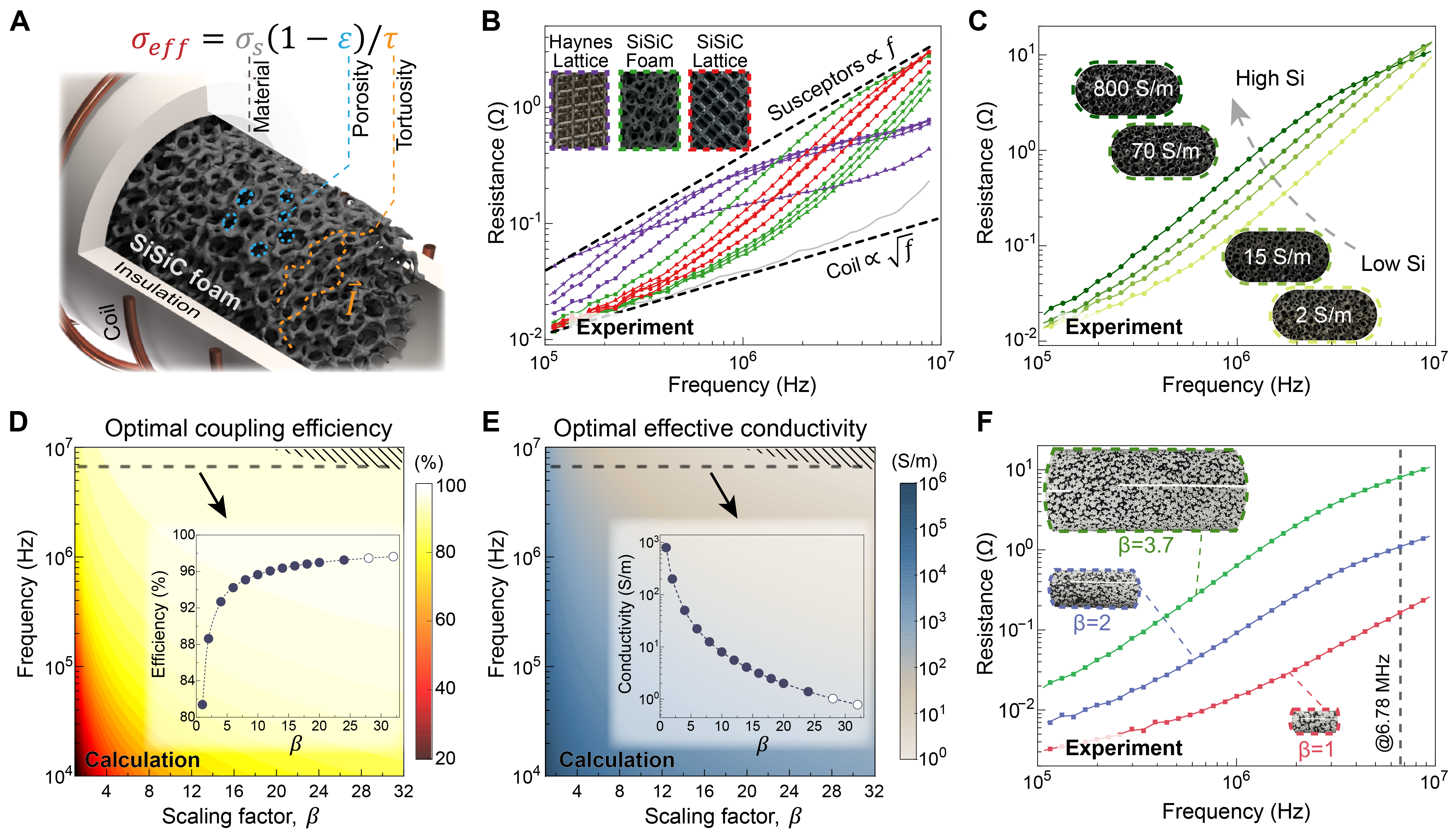}
\captionsetup{font=small}
\captionsetup{labelfont=bf}

\caption{\textbf{Effective conductivity properties of uniform metamaterial susceptors. (A)} Schematic of a metamaterial reactor comprising an open cell lattice susceptor that is inductively heated with a helical magnetic coil.  The susceptor effective electrical conductivity is specified by the Lemlich limit and is a function of material and geometric lattice parameters. \textbf{(B)} Experimental impedance measurements of different cylindrical metamaterial susceptors. Insets show images of foam and cubic lattice susceptors comprising Haynes superalloy and SiSiC. \textbf{(C)} Impedance characterization of 75 mm-diameter SiSiC foam susceptors with different silicon loadings, showing effective electrical conductivities ranging from 2 to 800 S/m. \textbf{(D)} Coupling efficiency versus $\beta$ at a fixed induction frequency of 6.78 MHz. As $\beta$ increases, efficiency approaches 100\%. \textbf{(E)} $\sigma_{\mathrm{eff}}$ tuning required for the scaling trend shown in \textbf{(D)}. \textbf{(F)} Impedance measurements of susceptors with $\beta$=1, 2, and 3.7, each with tuned $\sigma_{\mathrm{eff}}$ such that $f_{\mathrm{ideal}} \sim 6.78$ MHz. The insets are images of the catalyst-packed susceptors.}
\label{fig:fig_2}
\end{figure}

SiSiC, which will be the focus of the rest of this study, is  particularly ideal for scaled reactor implementation because of its chemical inertness, high temperature compatibility, low thermal expansion, and high thermal conductivity.\cite{Eom2013}  In addition, its material electrical conductivity can be tuned by adjusting its silicon content during manufacturing.\cite{Sangsuwan2001, Gianella2012}  To explore the extent of this tuning mechanism, we characterize cylindrical SiSiC foams (70 mm diameter) with consistent geometric parameters but made with different amounts of silicon loading.  The resulting resistance curves are shown in Fig. 2C, and curve fitting yields $\sigma_{\mathrm{eff}}$ values ranging from 2 to 800 S/m.  This ability to tune conductivity is particularly important for operating scaled, high frequency reactors.  To demonstrate, we fix $f$ to 6.78 MHz, which is a dedicated frequency in the electromagnetic industrial, scientific, and medical band,\cite{CFRPart18} and increase $\beta$.  We see in Fig. 2D that as $\beta$ increases, the calculated coupling efficiency increases and approaches 100\%.  The required $\sigma_{\mathrm{eff}}$ scales as approximately $1/\sqrt{\beta}$ (Fig. 2E).  This scaling regime can be captured experimentally, and we characterize a set of SiSiC foam susceptors representing different $\beta$ values (1, 2, and 3.7) with a tuned $\sigma_{\mathrm{eff}}$. The AC resistance curves (Fig. 2F) each show an  $f_{\mathrm{ideal}}$ at approximately 6.78 MHz, and fits of these curves with our AC resistance model yield effective conductivity values of 400, 201, and 70~S/m for $\beta=1$, 2, and 3.7, respectively. The AC resistance scales approximately as $\beta^3$ at this frequency, which matches our analytic theory.

\subsection{Total efficiency scaling}

A more comprehensive performance assessment of the total efficiency of scaled metamaterial reactors requires an analysis of heat utilization, heat loss, and temperature profiling within the reactor itself. We define total efficiency to be the ratio of energy used to heat the inlet gas outlet and drive the endothermic reaction divided by the electrical energy inputted into the power electronics. In particular, losses arise from heat loss to the environment, resistive loss in the induction coil, and inefficiencies in the power electronics. To perform this analysis, we consider the RWGS reaction as our model chemical system for study. The RWGS reaction is an endothermic reaction that converts CO$_2$ and H$_2$ into CO and H$_2$O, and it has been explored as a foundational reaction for the circular carbon economy due to its utilization of carbon dioxide.\cite{GonzalezCastano2021} We first perform experiments with lab scale reactors of different sizes (Fig.~3A) and match the reactor characteristics with multiphysics simulations (Fig.~3B) to produce an experimentally refined reactor model.  We then extrapolate this refined model to investigate the performance of larger reactors. The lab scale reactors used for experimental study utilize the susceptors shown in Fig.~2F .  The RWGS catalyst consists of 18\% wt K$_2$CO$_3$ on 1 mm-diameter mesoporous Al$_2$O$_3$ supports, and it is chosen because it features high activity at moderate temperatures ($\geq$430~$^\circ$C), 100\% selectivity for CO with no CH$_4$ byproduct, and excellent multi-day stability.\cite{Li2022, Tamakuwala2025}

\begin{figure}
\includegraphics[width=\textwidth]{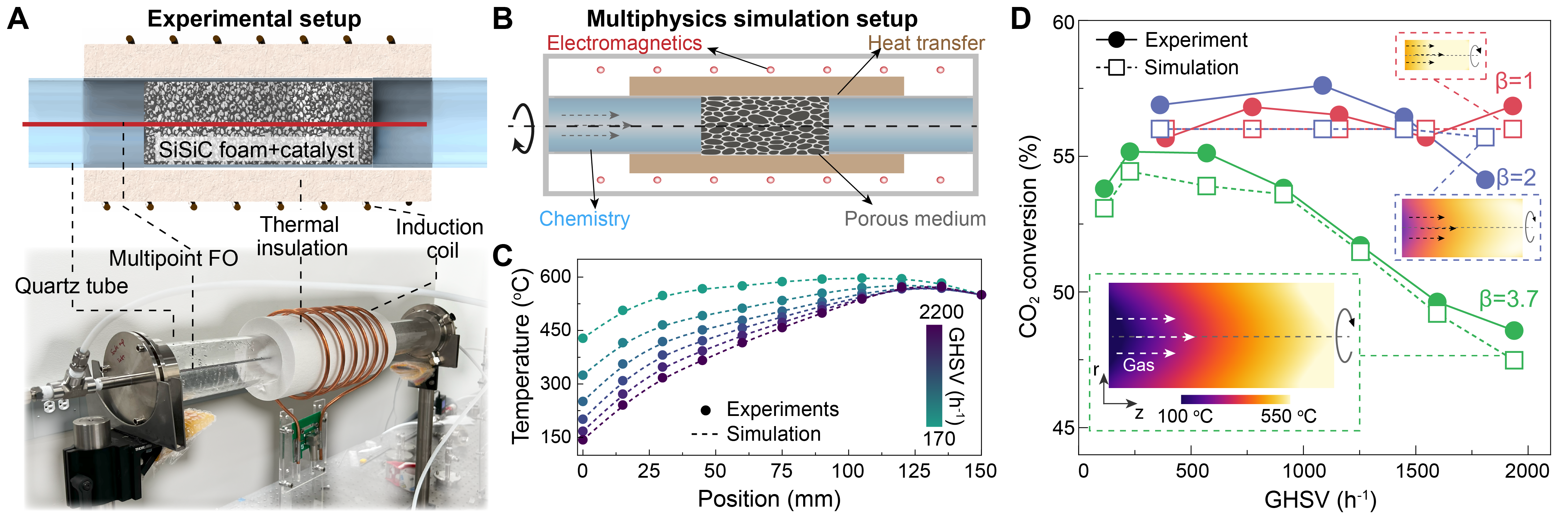}
\captionsetup{font=small}
\captionsetup{labelfont=bf}
\caption{\textbf{Experimentally refined multiphysics reactor model.}
\textbf{(A)} Cross sectional schematic (top) and image (bottom) of the experimental metamaterial reactor setup. 
\textbf{(B)} Schematic of our multiphysics modeling effort, which combines electromagnetics, heat transfer, fluid flow, and chemical reaction modalities. 
\textbf{(C)} Experimental axial temperature profiles across different GHSV conditions (solid circles) and corresponding simulated profiles (dashed lines), which are fitted using our multiphysics reactor model to the experimental profiles.
\textbf{(D)} CO$_2$ conversion as a function of GHSV for three reactor sizes, measured experimentally (solid circles) and simulated (open squares) using our fitted multiphysics model from (C). The experimental and simulated conversion agree well for all cases. The insets are simulated 2D temperature profiles of the reactors at high GHSV values.}
\label{fig:fig_3}
\end{figure}

A schematic and photograph of the experimental setup (Fig.~3A) shows a susceptor baffle that is packed with catalyst particles and housed within a quartz tube.  It is further surrounded by 25 mm-thick aluminosilicate thermal insulation and is heated using a helical induction coil. A multi-point fiber optic (FO) temperature sensor placed along the rotational symmetry axis is used to fix the outlet temperature to 550~$^\circ$C by providing feedback to the coil power supply, and it also simultaneously monitors the axial temperature profile within the reactor. A fiber optic sensor is used over a thermocouple due to its insensitivity to magnetic fields, ensuring accurate measurements during induction heating. We run RWGS reactions using this configuration for the three susceptor sizes and gas hourly space velocity (GHSV) values up to 1800 h$^{-1}$. Experimental axial temperature profiles are shown in Fig.~3C and experimental CO$_2$ conversion ($X_{CO_2}$) results are shown as solid dots in Fig.~3D. For the $\beta = $ 1 and 2 reactors, conversion values are near equilibrium values over the tested GHSV range. However, for the $\beta = 3.7$ reactor, conversion begins to deviate at higher GHSV values, suggesting a breakdown of plug flow-like reactor behavior.

Our experimentally refined reactor model uses COMSOL Multiphysics and considers the induction heating process, heat transfer, fluid flow, and reaction enthalpy in our model. To perform these multiphysics simulations, we first perform electromagnetics simulations to compute the induction heating profile within the susceptor. The electromagnetic properties of SiSiC are effectively independent of temperature, enabling the heating profile to be calculated independently of other physical parameters.  This profile is then inputted into a coupled heat transfer and reaction simulator in which heat transfer processes and heat consumption from the reaction are iteratively solved until self consistency.  For these simulations, we assume the packed bed reaction zone consists of a porous medium that is described using the Brinkman equation\cite{Durlofsky1987} and that has a void fraction of 0.5. The catalyst reaction kinetics and thermal conductivity of the reactor materials outside of the reaction zone are taken from our prior study.\cite{Lin2024}  Experimental model refinement is performed for the thermal conductivity of the reaction zone, which is extracted by fitting the model to the experimental axial temperature profiles (Fig.~3C) and is determined to be 11~W/(K$\cdot$m).

The calculated CO$_2$ conversion rates from our refined model are shown as open dots in Fig.~3D and agree well with our experimental data. The observed trends in CO$_2$ conversion can be further analyzed by calculating the full temperature profile within the reactor systems at high flow rates for different $\beta$, which are shown as insets in Fig.~3D. For $\beta = $1 and 2, the temperature profile supports small radial temperature gradients, indicating that the system operates near plug flow-like behavior due to the small dimensions of the system.  For $\beta = 3.7$, on the other hand, there are  significant radial temperature gradients that are responsible for the observed drop in CO$_2$ conversion. As such, while uniform metamaterial reactors attempt to manage radial temperature gradients by use of a volumetric heating profile and a high internal thermal conductivity, non-trivial gradients still persist for $\beta$ greater than 2.

To mitigate radial temperature gradients in scaled metamaterial reactors, we take advantage of one of the distinguishing features of metamaterial reactors: the ability to spatially tailor the heating profile through control of the local effective electrical conductivity. In particular, spatially varying effective conductivity profiles can be achieved by spatially modifying the local susceptor geometry and material composition (Fig.~2).  To demonstrate, we consider a metamaterial susceptor with a conductivity profile that follows $1/r^2$, which explicitly supports a volumetrically uniform heating profile when heated with a uniform axial magnetic field (Supplementary Section 3). In practice, it is not possible to specify this exact profile due to the divergence in conductivity at the rotational symmetry axis of the cylinder, and we instead consider the following effective conductivity profile:
\[
\sigma(r) =
\begin{cases}
\displaystyle \frac{\sigma_{\text{eff}}}{A} \cdot \left( \frac{R}{r} \right)^{2}, & \text{for } R/5 < r \leq R \\
\\
\displaystyle \frac{\sigma_{\text{eff}}\times 25}{A}, & \text{for } 0 \leq r \leq R/5
\end{cases}
\tag*{(1)}
\]
where $\sigma_\text{eff}$ is the corresponding effective conductivity chosen for the homogeneous susceptor and A is a hyperparameter that we set to 600.

To examine the impact of radial susceptor customization on reactor performance, we simulate the performance of scaled uniform and radially dependent metamaterial reactors utilizing our experimentally refined model for $\beta=32$. Plots of the the effective conductivity profiles, heating profiles, and radial temperature distribution for uniform and radially tailored reactors are shown in Figs.~4A and 4B, respectively. In both cases, the maximum temperature in the reactor is set to 550\textdegree C and the GHSV is chosen such that the conversion of carbon dioxide is 50\%. In addition, the frequency is chosen to be 100~kHz, which corresponds to the 95\% coupling efficiency contour line for homogeneous susceptors in Fig. 1B. 

\begin{figure}
\includegraphics[width=\textwidth]{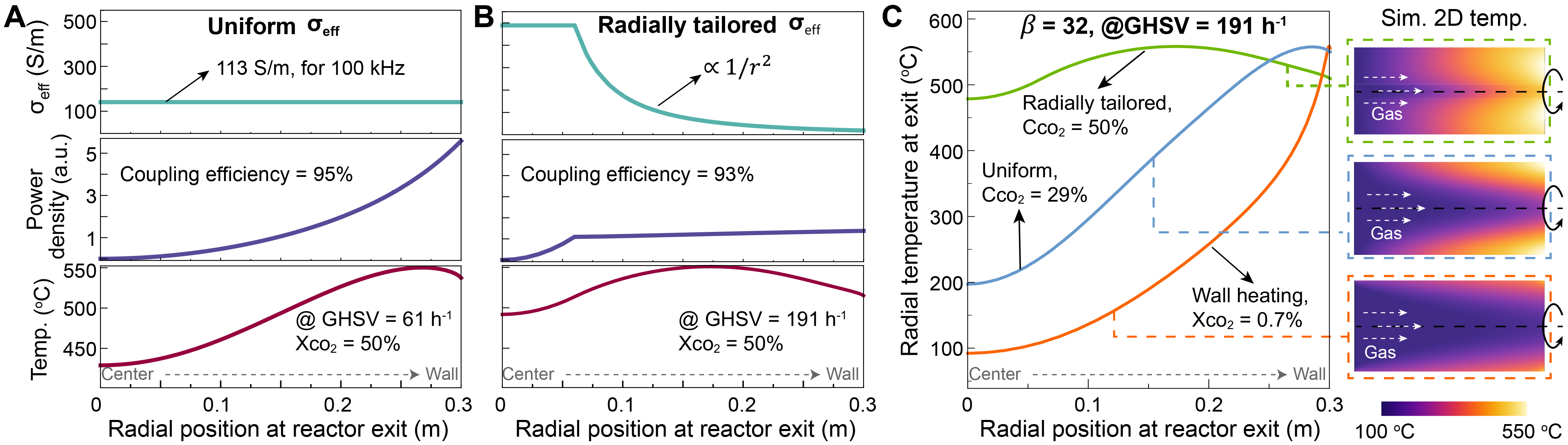}
\captionsetup{font=small}
\captionsetup{labelfont=bf}
\caption{\textbf{Radial tailoring of metamaterial susceptor effective conductivity.} 
\textbf{(A,B)} Effective conductivity profile (top panel), heat dissipation profile (middle panel) and radial temperature profile at the outlet (bottom panel) for a (A) uniform and (B) radially tailored metamaterial reactor.  Both reactors have a 0.6\,m diameter ($\beta=32$), operate with a 100 kHz frequency, have a maximum internal temperature of 550~$^\circ$C, and have a fixed CO$_2$ conversion of 50\%.  The uniform susceptor features a parabolic heating profile, which leads to strong radial temperature gradients and a GHSV of 61\,h$^{-1}$.  The radially tailored susceptor features a nearly uniform volumetric heating profile, which leads to reduced radial temperature gradients and a ${>}$3-fold increase in GHSV.
\textbf{(C)} Radial outlet temperature profiles, conversion, and axial temperature profiles of wall heated, uniform metamaterial, and radially tailored metamaterial reactors.  GHSV is fixed to 197\,h$^{-1}$ and $\beta=32$ in all cases.
}
\label{fig:fig_4}
\end{figure}

The heating profile for the uniform susceptor features a parabolic-like radial heating profile (middle panel, Fig. 4A), which is consistent with the volumetric heating profile in uniform cylindrical susceptors by helical coils at all scales. The underheated center region leads to significant radial temperature gradients at the outlet (bottom panel, Fig.4A) that limit the GHSV to 61~h$^{-1}$, indicating that relatively long residence times are required to achieve our desired conversion. Our radially tailored reactor, on the other hand, supports a nearly uniform heating profile (middle panel, Fig. 4B) while operating with a nearly identical coupling efficiency. The corresponding radial temperature profile at the outlet (bottom panel, Fig.4B) is much more uniform, though radial temperature gradients still persist due to heat loss through the thermal insulation and the lack of heating at the central region along the axis of the cylindrical reactor. These reactors support a GHSV of 197~h$^{-1}$, which is more than threefold higher than for the uniform case and which shows how radial tailoring can enable process intensification and the use of larger reactant flows.

To further evaluate the impact of radial tailoring on scaled reactor performance, we compare the capabilities of scaled uniform metamaterial reactors, radially tailored metamaterial reactors, and wall heated reactors under the same GHSV of 197~h$^{-1}$ and a maximum reactor temperature of 550\textdegree C. The wall heated reactor contains a SiSiC foam to enhance its internal thermal conductivity and improve heat spreading in the reactor. The radial temperature profiles at the outlet for all cases are shown in Fig. 4C. The wall heated reactor, which is included as a simple reference, exhibits extreme temperature gradients and displays nearly no conversion. The homogeneous metamaterial reactor also displays strong radial temperature gradients spanning hundreds of degrees that results in a conversion of 29\%.  The radially tailored reactor, on the other hand, has a conversion of 50\%, which is near the equilibrium conversion value of 55\%.  Cross sectional temperature plots of the reactors show that radial temperature gradients at all axial positions of the reactor are significantly reduced with the radial tailored metamaterial approach.

\begin{table}
\centering
\small
\caption*{\textbf{Table 1.} Simulation parameters for scale up analysis}
\begin{tabularx}{\textwidth}{>{\centering\arraybackslash}X >{\centering\arraybackslash}X >{\centering\arraybackslash}X >{\centering\arraybackslash}X >{\centering\arraybackslash}X >{\centering\arraybackslash}X >{\centering\arraybackslash}X}
\toprule
$\beta$ & Susceptor Diameter (m) & Frequency (MHz) & \multicolumn{2}{c}{Homogeneous susceptor} & \multicolumn{2}{c}{Radially tailored susceptor} \\
\cmidrule(lr){4-5} \cmidrule(lr){6-7}
& & & $\sigma_\mathrm{eff}$ (S/m) & Coupling Efficiency (\%) & $A$ for $\sigma(r)$ in Eq.~1 & Coupling Efficiency (\%) \\
\midrule
4  & 0.075  & 3      & 240.4     & 95  & N/A     & 93  \\
8  & 0.15   & 0.93   & 193.4     & 95  & 14.06  & 93 \\
16 & 0.3    & 0.3    & 150       & 95  & 11.25  & 93  \\
20 & 0.375  & 0.21   & 137.5     & 95  & 10.55  & 93  \\
24 & 0.45   & 0.16   & 127.7     & 95  & 10.13  & 93  \\
28 & 0.525  & 0.12   & 119.9     & 95  & 8.27   & 93  \\
32 & 0.6    & 0.1    & 113.4     & 95  & 7.2    & 93  \\
\bottomrule
\end{tabularx}
\label{tab:ghsv_table}
\end{table}

Finally, we perform a systematic analysis of scaled metamaterial reactor performance for $\beta = 4$--32. We consider uniform and radially tailored metamaterial reactors and wall heated reactors with SiSiC baffles. Our analysis can ultimately extend to larger $\beta$ without loss of generality, though we view moderate scale metamaterial reactors as representative of commercial scale systems due to the ability of metamaterial reactors to process intensify highly endothermic reactions and support enhanced product throughput. For all scales and reactor types, we set the induction frequency to follow the 95\% coupling efficiency contour line in Fig. 1B.  We assume that the efficiency of the power electronics itself is 95\%, which is consistent with reported efficiencies of wide bandgap-based amplifiers at megahertz frequencies and silicon-based amplifiers at kilohertz frequencies.  The effective conductivity values and profiles of the metamaterial susceptors are optimized for coupling efficiency for each $\beta$ value and susceptor type. A summary of chosen reactor parameters for different $\beta$ values are in Table~1.

We first use multiphysics simulations to determine the GHSV for each reactor type and size in which the maximum outlet temperature is 550$^\circ$C and the corresponding CO$_2$ conversion is 50\%. The resulting GHSV values as a function of $\beta$ are plotted in Fig.~5A. The uniform metamaterial reactor (blue curve, Fig.~5A) shows an order-of-magnitude GHSV enhancement for larger reactors compared to the wall heated reactor (orange curve, Fig.~5A), and the radially tailored reactor (green curve, Fig.~5A) yields an additional few-fold GHSV enhancement over the uniform reactor. These results demonstrate that while structured reactor concepts that combine wall heating with high thermal conductivity baffles have some efficacy, volumetric heated metamaterial susceptors provide a qualitatively different and better capability for $\beta \ge 8$.  Furthermore, while uniform and radially tailored reactors both support volumetric heating, the uniform volumetric heating provided by radial tailoring leads to enhanced performance for $\beta \ge 12$.

\begin{figure}
\centering
\includegraphics[width=0.75\textwidth]{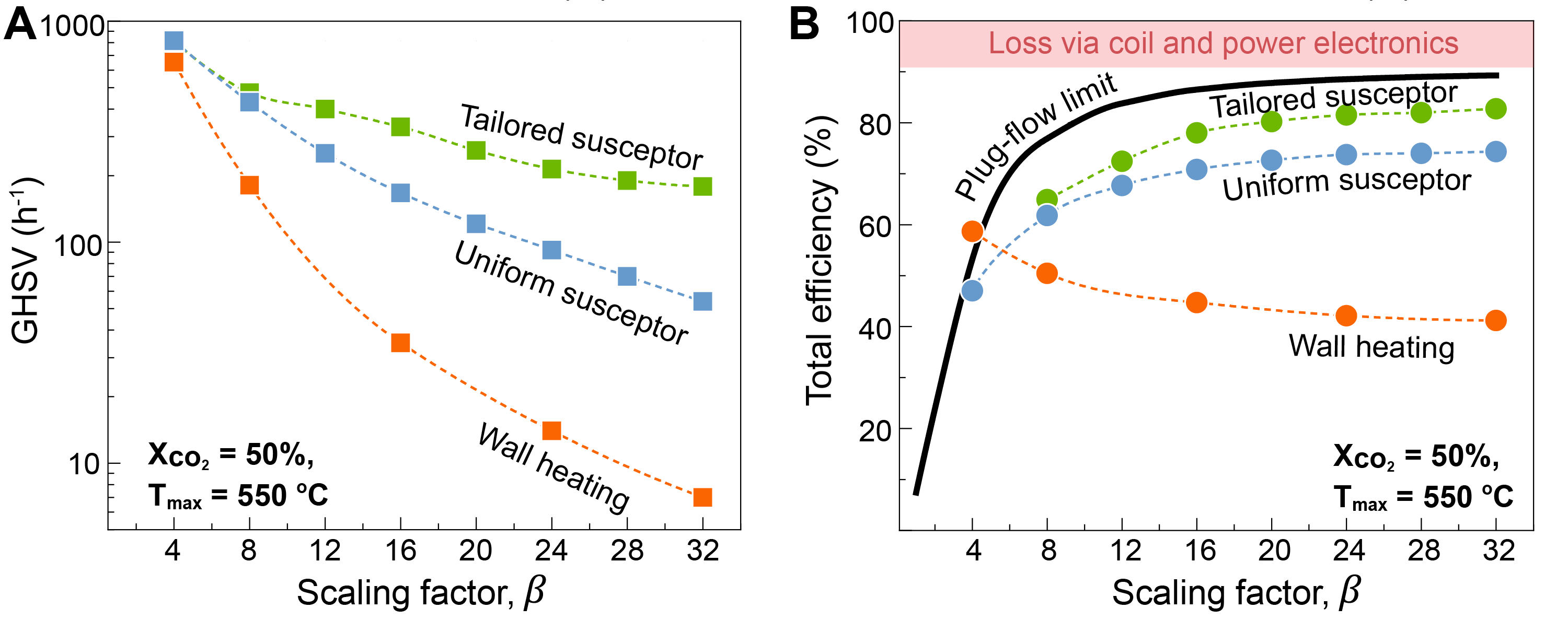}
\captionsetup{font=small}
\captionsetup{labelfont=bf}
\caption{\textbf{Process intensification and total efficiency metrics of metamaterial reactors for $\beta=4-32$.} 
\textbf{(A)} Plot of GHSV values required to reach 50\% CO$_2$ conversion for wall-heated, uniform metamaterial, and radially tailored metamaterial reactors of different scales.  The maximum outlet temperature is fixed to 550\,$^\circ$C.  
\textbf{(B)} Plot of total efficiencies of the three reactors in (A) as a function of $\beta$, together with the efficiencies from an ideal plug flow reactor.  The radially tailored metamaterial reactor closely follows the plug flow trend line and outperforms the other designs at scale.
}
\label{fig:fig_5}
\end{figure}

We next evaluate the total efficiency of the reactors as a function of $\beta$, assuming operation conditions from Fig.~5A, and the results are plotted in Fig.~5B. The black curve represents system efficiencies assuming ideal plug flow behavior, and it serves as a theoretical limit for fixed bed reactor operation. In this limit, the energy used to heat the gas and drive the endothermic reaction scales as $\beta^3$ and dominates in larger systems, eventually approaching the cap set by power electronics losses and coupling efficiency.\cite{Lin2024} The uniform metamaterial reactor has an overall efficiency that increases as a function of $\beta$, indicating that while its GHSV decreases with increasing reactor size, the scaling in energy used for gas heating and the reaction follows a power law that is greater than that for parasitic losses. The radially tailored metamaterial reactor has overall efficiencies that closely follow those predicted by the plug flow limit, highlighting the promise of spatially-engineered metamaterial susceptors in addressing scale-up challenges associated with nonuniform radial temperature distributions.

\section{Conclusions}

In summary, we present a scale up analysis of inductively heated metamaterial reactors. We utilize a combination of experimental and multiphysics analyses to elucidate how the induction frequency and susceptor conductivity in uniform metamaterial reactors can be tailored to support optimal performance in scaled systems.  While these reactors can be designed to feature high coupling efficiencies, radial temperature gradients ultimately limit process intensification capabilities and total efficiencies in scaled systems. We further propose the use of metamaterial susceptors with radially tailored effective electrical conductivity profiles that explicitly support uniform volumetric heating profiles.  We show that scaled reactor systems with these susceptors have plug flow-like chemical conversion capabilities and total efficiencies that are limited only by power electronics and coil coupling.

We envision multiple pathways for future study. One direction involves a more comprehensive investigation of spatially varying metamaterial profiles in scaled systems. While we discussed in this study effective conductivity profiles that support uniform volumetric heating,  more complex profiles can be inverse designed to enable further enhancements in temperature uniformity in scaled systems. These concepts can also extend to axial temperature customization to push process intensification to even more extreme limits. In another direction, a more detailed analysis and understanding of power electronics efficiency and scaling is required to more comprehensively understand the capabilities and performance limits of scaled electrified reactor systems. We also envision ample opportunities to understand how scaled metamaterial susceptors with desired properties can be effectively realized using industrially established manufacturing methods.

\begin{acknowledgement}

J. Fan, M. Kanan, and J. Rivas-Davila acknowledge support from the Department of Energy under agreement number DE-EE0011191.  A. Hofelmann and C. Lin acknowledge support from the Stanford Graduate Fellowship and D. Mantle acknowledges support from the TomKat Graduate Fellowship.

\end{acknowledgement}

\begin{suppinfo}

A separate PDF file is uploaded as supplementary information referred to in this manuscript. 

\begin{itemize}
  \item Filename: Supporting information: Scale up analysis of inductively heated metamaterial reactor
  
\end{itemize}

\end{suppinfo}

\bibliography{Main_text}

\providecommand{\latin}[1]{#1}
\makeatletter
\providecommand{\doi}
  {\begingroup\let\do\@makeother\dospecials
  \catcode`\{=1 \catcode`\}=2 \doi@aux}
\providecommand{\doi@aux}[1]{\endgroup\texttt{#1}}
\makeatother
\providecommand*\mcitethebibliography{\thebibliography}
\csname @ifundefined\endcsname{endmcitethebibliography}  {\let\endmcitethebibliography\endthebibliography}{}
\begin{mcitethebibliography}{44}
\providecommand*\natexlab[1]{#1}
\providecommand*\mciteSetBstSublistMode[1]{}
\providecommand*\mciteSetBstMaxWidthForm[2]{}
\providecommand*\mciteBstWouldAddEndPuncttrue
  {\def\EndOfBibitem{\unskip.}}
\providecommand*\mciteBstWouldAddEndPunctfalse
  {\let\EndOfBibitem\relax}
\providecommand*\mciteSetBstMidEndSepPunct[3]{}
\providecommand*\mciteSetBstSublistLabelBeginEnd[3]{}
\providecommand*\EndOfBibitem{}
\mciteSetBstSublistMode{f}
\mciteSetBstMaxWidthForm{subitem}{(\alph{mcitesubitemcount})}
\mciteSetBstSublistLabelBeginEnd
  {\mcitemaxwidthsubitemform\space}
  {\relax}
  {\relax}

\bibitem[Thiel and Stark(2021)Thiel, and Stark]{Thiel2021}
Thiel,~G.~P.; Stark,~A.~K. To decarbonize industry, we must decarbonize heat. \emph{Joule} \textbf{2021}, \emph{5}, 531--550\relax
\mciteBstWouldAddEndPuncttrue
\mciteSetBstMidEndSepPunct{\mcitedefaultmidpunct}
{\mcitedefaultendpunct}{\mcitedefaultseppunct}\relax
\EndOfBibitem
\bibitem[Van~Geem \latin{et~al.}(2019)Van~Geem, Galvita, and Marin]{VanGeem2019}
Van~Geem,~K.~M.; Galvita,~V.~V.; Marin,~G.~B. Making chemicals with electricity. \emph{Science} \textbf{2019}, \emph{364}, 734--735\relax
\mciteBstWouldAddEndPuncttrue
\mciteSetBstMidEndSepPunct{\mcitedefaultmidpunct}
{\mcitedefaultendpunct}{\mcitedefaultseppunct}\relax
\EndOfBibitem
\bibitem[Badakhsh \latin{et~al.}(2021)Badakhsh, Kwak, Lee, Jeong, Kim, Sohn, Nam, Yoon, Park, and Jo]{Badakhsh2021}
Badakhsh,~A.; Kwak,~Y.; Lee,~Y.-J.; Jeong,~H.; Kim,~Y.; Sohn,~H.; Nam,~S.~W.; Yoon,~C.~W.; Park,~C.~W.; Jo,~Y.~S. A compact catalytic foam reactor for decomposition of ammonia by the Joule-heating mechanism. \emph{Chemical Engineering Journal} \textbf{2021}, \emph{426}, 130802\relax
\mciteBstWouldAddEndPuncttrue
\mciteSetBstMidEndSepPunct{\mcitedefaultmidpunct}
{\mcitedefaultendpunct}{\mcitedefaultseppunct}\relax
\EndOfBibitem
\bibitem[Dong \latin{et~al.}(2023)Dong, Lele, Zhao, Li, Cheng, Wang, Cui, Guo, Brozena, Lin, Li, Xu, Qi, Kevrekidis, Mei, Pan, Liu, Ju, and Hu]{Dong2023}
Dong,~Q. \latin{et~al.}  Depolymerization of plastics by means of electrified spatiotemporal heating. \emph{Nature} \textbf{2023}, \emph{616}, 488--494\relax
\mciteBstWouldAddEndPuncttrue
\mciteSetBstMidEndSepPunct{\mcitedefaultmidpunct}
{\mcitedefaultendpunct}{\mcitedefaultseppunct}\relax
\EndOfBibitem
\bibitem[Wismann \latin{et~al.}(2019)Wismann, Engbæk, Vendelbo, Bendixen, Eriksen, Aasberg-Petersen, Frandsen, Chorkendorff, and Mortensen]{Wismann2019}
Wismann,~S.~T.; Engbæk,~J.~S.; Vendelbo,~S.~B.; Bendixen,~F.~B.; Eriksen,~W.~L.; Aasberg-Petersen,~K.; Frandsen,~C.; Chorkendorff,~I.; Mortensen,~P.~M. Electrified methane reforming: A compact approach to greener industrial hydrogen production. \emph{Science} \textbf{2019}, \emph{364}, 756--759\relax
\mciteBstWouldAddEndPuncttrue
\mciteSetBstMidEndSepPunct{\mcitedefaultmidpunct}
{\mcitedefaultendpunct}{\mcitedefaultseppunct}\relax
\EndOfBibitem
\bibitem[From \latin{et~al.}(2024)From, Partoont, Rautenbach, Østberg, Bentien, Aasberg-Petersen, and Mortensen]{From2024}
From,~T.~N.; Partoont,~B.; Rautenbach,~M.; Østberg,~M.; Bentien,~A.; Aasberg-Petersen,~K.; Mortensen,~P.~M. Electrified steam methane reforming of biogas for sustainable syngas manufacturing and next-generation of plant design: A pilot plant study. \emph{Chemical Engineering Journal} \textbf{2024}, \emph{479}, 147205\relax
\mciteBstWouldAddEndPuncttrue
\mciteSetBstMidEndSepPunct{\mcitedefaultmidpunct}
{\mcitedefaultendpunct}{\mcitedefaultseppunct}\relax
\EndOfBibitem
\bibitem[Hou \latin{et~al.}(2021)Hou, Zhen, Liu, Kuang, Gao, Luo, Deng, Chen, and Wang]{Hou2021}
Hou,~L.; Zhen,~X.; Liu,~L.; Kuang,~D.; Gao,~Y.; Luo,~H.; Deng,~L.; Chen,~C.; Wang,~S. Synthesis, thermal stability, magnetic properties, and microwave absorption applications of CoNi-C core-shell nanoparticles with tunable Co/Ni molar ratio. \emph{Results in Physics} \textbf{2021}, \emph{22}, 103893\relax
\mciteBstWouldAddEndPuncttrue
\mciteSetBstMidEndSepPunct{\mcitedefaultmidpunct}
{\mcitedefaultendpunct}{\mcitedefaultseppunct}\relax
\EndOfBibitem
\bibitem[Kappe \latin{et~al.}(2013)Kappe, Pieber, and Dallinger]{Kappe2013}
Kappe,~C.~O.; Pieber,~B.; Dallinger,~D. Microwave Effects in Organic Synthesis: Myth or Reality? \emph{Angewandte Chemie International Edition} \textbf{2013}, \emph{52}, 1088--1094\relax
\mciteBstWouldAddEndPuncttrue
\mciteSetBstMidEndSepPunct{\mcitedefaultmidpunct}
{\mcitedefaultendpunct}{\mcitedefaultseppunct}\relax
\EndOfBibitem
\bibitem[Baker-Fales \latin{et~al.}(2023)Baker-Fales, Chen, and Vlachos]{BakerFales2023}
Baker-Fales,~M.; Chen,~T.-Y.; Vlachos,~D.~G. Scale-up of microwave-assisted, continuous flow, liquid phase reactors: Application to 5-Hydroxymethylfurfural production. \emph{Chemical Engineering Journal} \textbf{2023}, \emph{454}, 139985\relax
\mciteBstWouldAddEndPuncttrue
\mciteSetBstMidEndSepPunct{\mcitedefaultmidpunct}
{\mcitedefaultendpunct}{\mcitedefaultseppunct}\relax
\EndOfBibitem
\bibitem[Sundaramoorthy \latin{et~al.}(2025)Sundaramoorthy, Lobo, and Vlachos]{Sundaramoorthy2025}
Sundaramoorthy,~A.~S.; Lobo,~R.~F.; Vlachos,~D.~G. Coarse-Grained Models for Scale-Up of Structured Reactors. \emph{Ind.~Eng.~Chem.~Res.} \textbf{2025}, \emph{64}, 14110--14121\relax
\mciteBstWouldAddEndPuncttrue
\mciteSetBstMidEndSepPunct{\mcitedefaultmidpunct}
{\mcitedefaultendpunct}{\mcitedefaultseppunct}\relax
\EndOfBibitem
\bibitem[Morais \latin{et~al.}(2023)Morais, Delikonstantis, Scapinello, Smith, Stefanidis, and Bogaerts]{Morais2023}
Morais,~E.; Delikonstantis,~E.; Scapinello,~M.; Smith,~G.; Stefanidis,~G.~D.; Bogaerts,~A. Methane coupling in nanosecond pulsed plasmas: Correlation between temperature and pressure and effects on product selectivity. \emph{Chem.~Eng.~J.} \textbf{2023}, \emph{462}, 142227\relax
\mciteBstWouldAddEndPuncttrue
\mciteSetBstMidEndSepPunct{\mcitedefaultmidpunct}
{\mcitedefaultendpunct}{\mcitedefaultseppunct}\relax
\EndOfBibitem
\bibitem[Akande and Lee(2022)Akande, and Lee]{Akande2022}
Akande,~O.; Lee,~B. Plasma steam methane reforming (PSMR) using a microwave torch for commercial-scale distributed hydrogen production. \emph{Int.~J.~Hydrog.~Energy} \textbf{2022}, \emph{47}, 2874--2884\relax
\mciteBstWouldAddEndPuncttrue
\mciteSetBstMidEndSepPunct{\mcitedefaultmidpunct}
{\mcitedefaultendpunct}{\mcitedefaultseppunct}\relax
\EndOfBibitem
\bibitem[Akay \latin{et~al.}(2020)Akay, Zhang, Al-Harrasi, and Sankaran]{Akay2020}
Akay,~G.; Zhang,~K.; Al-Harrasi,~W. S.~S.; Sankaran,~R.~M. Catalytic Plasma Fischer–Tropsch Synthesis Using Hierarchically Connected Porous Co/SiO\textsubscript{2} Catalysts Prepared by Microwave-Induced Co-assembly. \emph{Ind.~Eng.~Chem.~Res.} \textbf{2020}, \emph{59}, 12013--12027\relax
\mciteBstWouldAddEndPuncttrue
\mciteSetBstMidEndSepPunct{\mcitedefaultmidpunct}
{\mcitedefaultendpunct}{\mcitedefaultseppunct}\relax
\EndOfBibitem
\bibitem[Mohamed \latin{et~al.}(2025)Mohamed, Kumarachari, Bukke, Neerugatti, Mekasha, and Bandarupalle]{Mohamed2025}
Mohamed,~R. Y.~A.; Kumarachari,~R.~K.; Bukke,~S. P.~N.; Neerugatti,~D.; Mekasha,~Y.~T.; Bandarupalle,~K. Plasma catalysis for sustainable industry: lab-scale studies and pathways to upscaling. \emph{Discov.~Appl.~Sci.} \textbf{2025}, \emph{7}, 271\relax
\mciteBstWouldAddEndPuncttrue
\mciteSetBstMidEndSepPunct{\mcitedefaultmidpunct}
{\mcitedefaultendpunct}{\mcitedefaultseppunct}\relax
\EndOfBibitem
\bibitem[Ceylan \latin{et~al.}(2008)Ceylan, Friese, Lammel, Mazac, and Kirschning]{Ceylan2008}
Ceylan,~S.; Friese,~C.; Lammel,~C.; Mazac,~K.; Kirschning,~A. Inductive Heating for Organic Synthesis by Using Functionalized Magnetic Nanoparticles Inside Microreactors. \emph{Angewandte Chemie International Edition} \textbf{2008}, \emph{47}, 8950--8953\relax
\mciteBstWouldAddEndPuncttrue
\mciteSetBstMidEndSepPunct{\mcitedefaultmidpunct}
{\mcitedefaultendpunct}{\mcitedefaultseppunct}\relax
\EndOfBibitem
\bibitem[Scarfiello \latin{et~al.}(2021)Scarfiello, Bellusci, Pilloni, Pietragiacomi, La~Barbera, and Varsano]{Scarfiello2021}
Scarfiello,~C.; Bellusci,~M.; Pilloni,~L.; Pietragiacomi,~D.; La~Barbera,~A.; Varsano,~F. Supported catalysts for induction-heated steam reforming of methane. \emph{International Journal of Hydrogen Energy} \textbf{2021}, \emph{46}, 134--145\relax
\mciteBstWouldAddEndPuncttrue
\mciteSetBstMidEndSepPunct{\mcitedefaultmidpunct}
{\mcitedefaultendpunct}{\mcitedefaultseppunct}\relax
\EndOfBibitem
\bibitem[Faure \latin{et~al.}(2021)Faure, Kale, Mille, Cayez, Ourlin, Soulantica, Carrey, and Chaudret]{Faure2021}
Faure,~S.; Kale,~S.~S.; Mille,~N.; Cayez,~S.; Ourlin,~T.; Soulantica,~K.; Carrey,~J.; Chaudret,~B. Improving energy efficiency of magnetic CO2 methanation by modifying coil design, heating agents, and by using eddy currents as the complementary heating source. \emph{Journal of Applied Physics} \textbf{2021}, \emph{129}, 044901\relax
\mciteBstWouldAddEndPuncttrue
\mciteSetBstMidEndSepPunct{\mcitedefaultmidpunct}
{\mcitedefaultendpunct}{\mcitedefaultseppunct}\relax
\EndOfBibitem
\bibitem[Teel \latin{et~al.}(2025)Teel, Craps, Mercado, Ciesielski, and Shimpalee]{Teel2025}
Teel,~H.; Craps,~M.; Mercado,~H.-C.; Ciesielski,~P.; Shimpalee,~S. Numerical simulation for the design of induction heating based radio frequency reactor for ethylene production. \emph{Chemical Engineering Research and Design} \textbf{2025}, \emph{219}, 511--521\relax
\mciteBstWouldAddEndPuncttrue
\mciteSetBstMidEndSepPunct{\mcitedefaultmidpunct}
{\mcitedefaultendpunct}{\mcitedefaultseppunct}\relax
\EndOfBibitem
\bibitem[Yan \latin{et~al.}(2024)Yan, Li, Pan, Shi, Xie, Liu, and Liu]{Yan2024}
Yan,~Y.; Li,~N.; Pan,~Y.; Shi,~L.; Xie,~G.; Liu,~Z.; Liu,~Q. Hydrogen-rich syngas production from tobacco stem pyrolysis in an electromagnetic induction heating fluidized bed reactor. \emph{International Journal of Hydrogen Energy} \textbf{2024}, \emph{68}, 1271--1280\relax
\mciteBstWouldAddEndPuncttrue
\mciteSetBstMidEndSepPunct{\mcitedefaultmidpunct}
{\mcitedefaultendpunct}{\mcitedefaultseppunct}\relax
\EndOfBibitem
\bibitem[Kim \latin{et~al.}(2023)Kim, Lee, and Lee]{Kim2023}
Kim,~Y.~T.; Lee,~J.-J.; Lee,~J. Electricity-driven reactors that promote thermochemical catalytic reactions via joule and induction heating. \emph{Chemical Engineering Journal} \textbf{2023}, \emph{470}, 144333\relax
\mciteBstWouldAddEndPuncttrue
\mciteSetBstMidEndSepPunct{\mcitedefaultmidpunct}
{\mcitedefaultendpunct}{\mcitedefaultseppunct}\relax
\EndOfBibitem
\bibitem[Wang \latin{et~al.}(2024)Wang, Otor, Rivera-Castro, and Hicks]{Wang2024}
Wang,~N.; Otor,~H.~O.; Rivera-Castro,~G.; Hicks,~J.~C. Plasma Catalysis for Hydrogen Production: A Bright Future for Decarbonization. \emph{ACS Catal.} \textbf{2024}, \emph{14}, 6749--6798\relax
\mciteBstWouldAddEndPuncttrue
\mciteSetBstMidEndSepPunct{\mcitedefaultmidpunct}
{\mcitedefaultendpunct}{\mcitedefaultseppunct}\relax
\EndOfBibitem
\bibitem[Lucía \latin{et~al.}(2014)Lucía, Maussion, Dede, and Burdío]{Lucia2014}
Lucía,~O.; Maussion,~P.; Dede,~E.~J.; Burdío,~J.~M. Induction Heating Technology and Its Applications: Past Developments, Current Technology, and Future Challenges. \emph{IEEE Transactions on Industrial Electronics} \textbf{2014}, \emph{61}, 2509--2520\relax
\mciteBstWouldAddEndPuncttrue
\mciteSetBstMidEndSepPunct{\mcitedefaultmidpunct}
{\mcitedefaultendpunct}{\mcitedefaultseppunct}\relax
\EndOfBibitem
\bibitem[Zhao \latin{et~al.}(2017)Zhao, Wu, He, Xiao, and Webley]{Zhao2017}
Zhao,~Q.; Wu,~F.; He,~Y.; Xiao,~P.; Webley,~P.~A. Impact of operating parameters on CO2 capture using carbon monolith by Electrical Swing Adsorption technology (ESA). \emph{Chemical Engineering Journal} \textbf{2017}, \emph{327}, 441--453\relax
\mciteBstWouldAddEndPuncttrue
\mciteSetBstMidEndSepPunct{\mcitedefaultmidpunct}
{\mcitedefaultendpunct}{\mcitedefaultseppunct}\relax
\EndOfBibitem
\bibitem[Deng \latin{et~al.}(2025)Deng, Eddy, Wyss, Tiwary, and Tour]{Deng2025}
Deng,~B.; Eddy,~L.; Wyss,~K.~M.; Tiwary,~C.~S.; Tour,~J.~M. Flash Joule heating for synthesis, upcycling and remediation. \emph{Nature Reviews Clean Technology} \textbf{2025}, \emph{1}, 32--54\relax
\mciteBstWouldAddEndPuncttrue
\mciteSetBstMidEndSepPunct{\mcitedefaultmidpunct}
{\mcitedefaultendpunct}{\mcitedefaultseppunct}\relax
\EndOfBibitem
\bibitem[Mallapragada \latin{et~al.}(2023)Mallapragada, Dvorkin, Modestino, Esposito, Smith, Hodge, Harold, Donnelly, Nuz, Bloomquist, Baker, Grabow, Yan, Rajput, Hartman, Biddinger, Aydil, and Taylor]{Mallapragada2023}
Mallapragada,~D.~S. \latin{et~al.}  Decarbonization of the chemical industry through electrification: Barriers and opportunities. \emph{Joule} \textbf{2023}, \emph{7}, 23--41\relax
\mciteBstWouldAddEndPuncttrue
\mciteSetBstMidEndSepPunct{\mcitedefaultmidpunct}
{\mcitedefaultendpunct}{\mcitedefaultseppunct}\relax
\EndOfBibitem
\bibitem[Noble \latin{et~al.}(2024)Noble, Bending, and Hill]{Noble2024}
Noble,~J. P.~P.; Bending,~S.~J.; Hill,~A.~K. Radiofrequency Induction Heating for Green Chemicals Manufacture: A Systematic Model of Energy Losses and a Scale-Up Case-Study. \emph{ACS Engineering Au} \textbf{2024}, \emph{4}, 450--463\relax
\mciteBstWouldAddEndPuncttrue
\mciteSetBstMidEndSepPunct{\mcitedefaultmidpunct}
{\mcitedefaultendpunct}{\mcitedefaultseppunct}\relax
\EndOfBibitem
\bibitem[Rudnev \latin{et~al.}(2017)Rudnev, Loveless, and Cook]{Rudnev2017}
Rudnev,~V.; Loveless,~D.; Cook,~R.~L. \emph{Handbook of Induction Heating}, 2nd ed.; CRC Press: Boca Raton, 2017; p 772\relax
\mciteBstWouldAddEndPuncttrue
\mciteSetBstMidEndSepPunct{\mcitedefaultmidpunct}
{\mcitedefaultendpunct}{\mcitedefaultseppunct}\relax
\EndOfBibitem
\bibitem[Ceylan \latin{et~al.}(2011)Ceylan, Coutable, Wegner, and Kirschning]{Ceylan2011}
Ceylan,~S.; Coutable,~L.; Wegner,~J.; Kirschning,~A. Inductive Heating with Magnetic Materials inside Flow Reactors. \emph{Chemistry – A European Journal} \textbf{2011}, \emph{17}, 1884--1893\relax
\mciteBstWouldAddEndPuncttrue
\mciteSetBstMidEndSepPunct{\mcitedefaultmidpunct}
{\mcitedefaultendpunct}{\mcitedefaultseppunct}\relax
\EndOfBibitem
\bibitem[Wang \latin{et~al.}(2019)Wang, Tuci, Duong-Viet, Liu, and Rossin]{Wang2019}
Wang,~W.; Tuci,~G.; Duong-Viet,~C.; Liu,~Y.; Rossin,~A. Induction Heating: An Enabling Technology for the Heat Management in Catalytic Processes. \emph{ACS Catalysis} \textbf{2019}, \emph{9}, 7921--7935\relax
\mciteBstWouldAddEndPuncttrue
\mciteSetBstMidEndSepPunct{\mcitedefaultmidpunct}
{\mcitedefaultendpunct}{\mcitedefaultseppunct}\relax
\EndOfBibitem
\bibitem[Lin \latin{et~al.}(2024)Lin, Wan, Ru, Cremers, Mohapatra, Mantle, Tamakuwala, Hoffelmann, Kanan, Rivas-Davila, and Fan]{Lin2024}
Lin,~C.~H.; Wan,~C.; Ru,~Z.; Cremers,~C.; Mohapatra,~P.; Mantle,~D.~L.; Tamakuwala,~K.; Hoffelmann,~A.~B.; Kanan,~M.~W.; Rivas-Davila,~J.; Fan,~J.~A. Electrified thermochemical reaction systems with high-frequency metamaterial reactors. \emph{Joule} \textbf{2024}, \emph{8}, 2938--2949\relax
\mciteBstWouldAddEndPuncttrue
\mciteSetBstMidEndSepPunct{\mcitedefaultmidpunct}
{\mcitedefaultendpunct}{\mcitedefaultseppunct}\relax
\EndOfBibitem
\bibitem[Ambrosetti \latin{et~al.}(2020)Ambrosetti, Bracconi, Maestri, Groppi, and Tronconi]{Ambrosetti2020}
Ambrosetti,~M.; Bracconi,~M.; Maestri,~M.; Groppi,~G.; Tronconi,~E. Packed foams for the intensification of catalytic processes: assessment of packing efficiency and pressure drop using a combined experimental and numerical approach. \emph{Chemical Engineering Journal} \textbf{2020}, \emph{382}, 122801\relax
\mciteBstWouldAddEndPuncttrue
\mciteSetBstMidEndSepPunct{\mcitedefaultmidpunct}
{\mcitedefaultendpunct}{\mcitedefaultseppunct}\relax
\EndOfBibitem
\bibitem[Visconti \latin{et~al.}(2016)Visconti, Groppi, and Tronconi]{Visconti2016}
Visconti,~C.~G.; Groppi,~G.; Tronconi,~E. Highly conductive “packed foams”: A new concept for the intensification of strongly endo- and exo-thermic catalytic processes in compact tubular reactors. \emph{Catalysis Today} \textbf{2016}, \emph{273}, 178--186, 5th International Conference on Structured Catalysts and Reactors, ICOSCAR-5, Donostia-San Sebastián, Spain, 22–24 June, 2016\relax
\mciteBstWouldAddEndPuncttrue
\mciteSetBstMidEndSepPunct{\mcitedefaultmidpunct}
{\mcitedefaultendpunct}{\mcitedefaultseppunct}\relax
\EndOfBibitem
\bibitem[Eigenberger and Ruppel(2012)Eigenberger, and Ruppel]{Eigenberger2012}
Eigenberger,~G.; Ruppel,~W. \emph{Ullmann's Encyclopedia of Industrial Chemistry}; John Wiley \& Sons, Ltd, 2012\relax
\mciteBstWouldAddEndPuncttrue
\mciteSetBstMidEndSepPunct{\mcitedefaultmidpunct}
{\mcitedefaultendpunct}{\mcitedefaultseppunct}\relax
\EndOfBibitem
\bibitem[Kapteijn and Moulijn(2022)Kapteijn, and Moulijn]{Kapteijn2022}
Kapteijn,~F.; Moulijn,~J.~A. Structured catalysts and reactors – Perspectives for demanding applications. \emph{Catalysis Today} \textbf{2022}, \emph{383}, 5--14, SI: STRUCTURED CATALYSTS\relax
\mciteBstWouldAddEndPuncttrue
\mciteSetBstMidEndSepPunct{\mcitedefaultmidpunct}
{\mcitedefaultendpunct}{\mcitedefaultseppunct}\relax
\EndOfBibitem
\bibitem[Lemlich(1978)]{Lemlich1978}
Lemlich,~R. A theory for the limiting conductivity of polyhedral foam at low density. \emph{Journal of Colloid and Interface Science} \textbf{1978}, \emph{64}, 107--110\relax
\mciteBstWouldAddEndPuncttrue
\mciteSetBstMidEndSepPunct{\mcitedefaultmidpunct}
{\mcitedefaultendpunct}{\mcitedefaultseppunct}\relax
\EndOfBibitem
\bibitem[Eom \latin{et~al.}(2013)Eom, Kim, and Raju]{Eom2013}
Eom,~J.-H.; Kim,~Y.-W.; Raju,~S. Processing and properties of macroporous silicon carbide ceramics: A review. \emph{Journal of Asian Ceramic Societies} \textbf{2013}, \emph{1}, 220--242\relax
\mciteBstWouldAddEndPuncttrue
\mciteSetBstMidEndSepPunct{\mcitedefaultmidpunct}
{\mcitedefaultendpunct}{\mcitedefaultseppunct}\relax
\EndOfBibitem
\bibitem[Sangsuwan \latin{et~al.}(2001)Sangsuwan, Orejas, Gatica, Tewari, and Singh]{Sangsuwan2001}
Sangsuwan,~P.; Orejas,~J.~A.; Gatica,~J.~E.; Tewari,~S.~N.; Singh,~M. Reaction-Bonded Silicon Carbide by Reactive Infiltration. \emph{Industrial \& Engineering Chemistry Research} \textbf{2001}, \emph{40}, 5191--5198\relax
\mciteBstWouldAddEndPuncttrue
\mciteSetBstMidEndSepPunct{\mcitedefaultmidpunct}
{\mcitedefaultendpunct}{\mcitedefaultseppunct}\relax
\EndOfBibitem
\bibitem[Gianella \latin{et~al.}(2012)Gianella, Gaia, and Ortona]{Gianella2012}
Gianella,~S.; Gaia,~D.; Ortona,~A. High Temperature Applications of SiSiC Cellular Ceramics. \emph{Advanced Engineering Materials} \textbf{2012}, \emph{14}, 1074--1081\relax
\mciteBstWouldAddEndPuncttrue
\mciteSetBstMidEndSepPunct{\mcitedefaultmidpunct}
{\mcitedefaultendpunct}{\mcitedefaultseppunct}\relax
\EndOfBibitem
\bibitem[CFR(2025)]{CFRPart18}
47 CFR Part 18 -- Industrial, Scientific, and Medical Equipment. \url{https://www.ecfr.gov/current/title-47/part-18}, 2025; Accessed: 2025-07-25\relax
\mciteBstWouldAddEndPuncttrue
\mciteSetBstMidEndSepPunct{\mcitedefaultmidpunct}
{\mcitedefaultendpunct}{\mcitedefaultseppunct}\relax
\EndOfBibitem
\bibitem[González-Castaño \latin{et~al.}(2021)González-Castaño, Dorneanu, and Arellano-García]{GonzalezCastano2021}
González-Castaño,~M.; Dorneanu,~B.; Arellano-García,~H. The reverse water gas shift reaction: a process systems engineering perspective. \emph{Reaction Chemistry \& Engineering} \textbf{2021}, \emph{6}, 954--976\relax
\mciteBstWouldAddEndPuncttrue
\mciteSetBstMidEndSepPunct{\mcitedefaultmidpunct}
{\mcitedefaultendpunct}{\mcitedefaultseppunct}\relax
\EndOfBibitem
\bibitem[Li \latin{et~al.}(2022)Li, Frankhouser, and Kanan]{Li2022}
Li,~C.~S.; Frankhouser,~A.~D.; Kanan,~M.~W. Carbonate-catalyzed reverse water-gas shift to produce gas fermentation feedstocks for renewable liquid fuel synthesis. \emph{Cell Reports Physical Science} \textbf{2022}, \emph{3}, 101021\relax
\mciteBstWouldAddEndPuncttrue
\mciteSetBstMidEndSepPunct{\mcitedefaultmidpunct}
{\mcitedefaultendpunct}{\mcitedefaultseppunct}\relax
\EndOfBibitem
\bibitem[Tamakuwala \latin{et~al.}(2025)Tamakuwala, Kennedy, Li, Mutz, Boller, Bare, and Kanan]{Tamakuwala2025}
Tamakuwala,~K.~N.; Kennedy,~R.~P.; Li,~C.~S.; Mutz,~B.; Boller,~P.; Bare,~S.~R.; Kanan,~M.~W. Intermediate-Temperature Reverse Water–Gas Shift under Process-Relevant Conditions Catalyzed by Dispersed Alkali Carbonates. \emph{JACS Au} \textbf{2025}, \emph{5}, 1083--1089\relax
\mciteBstWouldAddEndPuncttrue
\mciteSetBstMidEndSepPunct{\mcitedefaultmidpunct}
{\mcitedefaultendpunct}{\mcitedefaultseppunct}\relax
\EndOfBibitem
\bibitem[Durlofsky and Brady(1987)Durlofsky, and Brady]{Durlofsky1987}
Durlofsky,~L.; Brady,~J.~F. Analysis of the Brinkman equation as a model for flow in porous media. \emph{The Physics of Fluids} \textbf{1987}, \emph{30}, 3329--3341\relax
\mciteBstWouldAddEndPuncttrue
\mciteSetBstMidEndSepPunct{\mcitedefaultmidpunct}
{\mcitedefaultendpunct}{\mcitedefaultseppunct}\relax
\EndOfBibitem
\end{mcitethebibliography}

\clearpage
\begin{center}
    \LARGE \textbf{Supporting Information}
\end{center}

\subsection*{Section S1. Calculation of power dissipation and coupling efficiency}

As discussed in the main text, the foam can be treated as a homogeneous medium with an effective conductivity $\sigma_{\text{eff}}$ that depends on the pore geometry and material composition. Following this effective medium approximation, we can analytically compute the impedance and power dissipation in an inductively heated non-magnetic susceptor. 

We consider a susceptor with radius $a$ in the presence of an axially oriented magnetic field with magnitude $B_0$. We assume the susceptor and coil both have a total length of $L$. The magnetic field profile is described by the Helmholtz equation:

\begin{equation}
r^2 \frac{\partial^2 B_e}{\partial r^2} + r\frac{\partial B_e}{\partial r} + k^2 r^2 B_e = -k^2 r^2 B_0 \quad (r \leq R)
\tag{S1}
\end{equation}

The induced magnetic field $B_e$ is expressed in cylindrical coordinates:

\begin{equation}
B_z = B_e(r)\hat{z}, \quad \frac{\partial}{\partial \phi} = \frac{\partial}{\partial z} = 0
\tag{S2}
\end{equation}

By solving the Helmholtz equation and applying Ampère's law, the eddy current density within the susceptor is:

\begin{equation}
\mathbf{J}(r) = \frac{k B_z}{\mu_0} \left[ \frac{J_1(kr)}{J_0(kR)} \right] \hat{\phi} \quad (r \leq a)
\tag{S3}
\end{equation}

where $J_i$ is the Bessel function of the first kind of order $i$. The total power dissipation $P_{\text{diss}}$ is obtained by integrating the square of the current density over the volume of the cylinder:

\begin{equation}
P_{\text{diss}} = \frac{1}{2 \sigma_{\text{eff}}} \int |J|^2 dV 
= \frac{2 \pi\omega B_z^2 L R^2}{\mu_0} 
\left[ \frac{J_2((1+i)R/\delta)}{J_0((1+i)R/\delta)} \right]
\tag{S4}
\end{equation}

where $\delta$ is the skin depth in the susceptor.

We calculated the axial magnetic field $B_z$ in the susceptor using the Biot–Savart law, assuming a total coil length of $2L_c$ and coil radius $R_c$, and the final result is expressed as:

\begin{equation}
B_z = \frac{\mu_0 N I}{4L} \left( \frac{(z + L_c)}{\sqrt{(z + L_c)^2 + R_c^2}} + \frac{(L_c - z)}{\sqrt{(L_c - z)^2 + R_c^2}} \right)
\tag{S5}
\end{equation}

where $N$ is the number of coil turns and $I$ is the RMS current in the coil.

Using Eqs. (S4) and (S5), the impedance of the susceptor, $R_{\text{sus}}$, is:

\begin{equation}
R_{\text{sus}} = \frac{P_{\text{diss}}}{I^2}
\tag{S6}
\end{equation}

For our copper induction coil, we use the improved Dowell’s method [cite] that accounts for skin and proximity effects to approximate the AC resistance:

\begin{equation}
R_{\text{coil}} = \frac{\pi^{3/4} d N^{3/2}}{\sigma_c \delta_c \sqrt{2a_c L_c}}
\tag{S7}
\end{equation}

where $\sigma_c$ is the electrical conductivity of copper, $a_c$ is the radius of the copper tube, and $\delta_c$ is the skin depth of the copper.

The coupling efficiency, away from the coil’s resonant frequency, is given by:

\begin{equation}
\eta_{\text{coupling}} = \frac{R_{\text{sus}}}{R_{\text{sus}} + R_{\text{coil}}}
\tag{S8}
\end{equation}

In this study, we fixed the number of coil turns ($N$) at 7 across all reactor scales. The inner diameter of the coil was set to be 10 mm larger than the insulation diameter. To ensure consistent heating, we aligned the middle five turns of the coil with the length of the susceptor, allowing one additional turn to extend above and below the susceptor ends. This configuration helps maintain a fairly uniform magnetic field across the susceptor region. The diameter of the copper wire was scaled proportionally to the square root of $\beta$. Based on this geometry and the preceding derivation, the coupling efficiencies for all reactor scales were calculated and plotted in Figs. 1 and 2 of the main text.

\subsection*{Section S2. Evolution of radial temperature gradients during scale-up of homogeneous susceptors}

As discussed in the main text, by co-designing the effective electrical conductivity of the susceptor and the induction frequency, the volumetric heating profile and coupling efficiency can be optimized for a given reactor diameter. Using the optimized parameters (summarized in Table~1), we analyzed scale-up performance from $\beta = 1$ to 32. Our target was to find the maximum GHSV that supports 50\% conversion while maintaining an exit temperature no greater than 550\,°C. The resulting radial temperature profiles for different reactor sizes are shown in Fig.~S1.

\begin{center}
    \includegraphics[width=\textwidth]{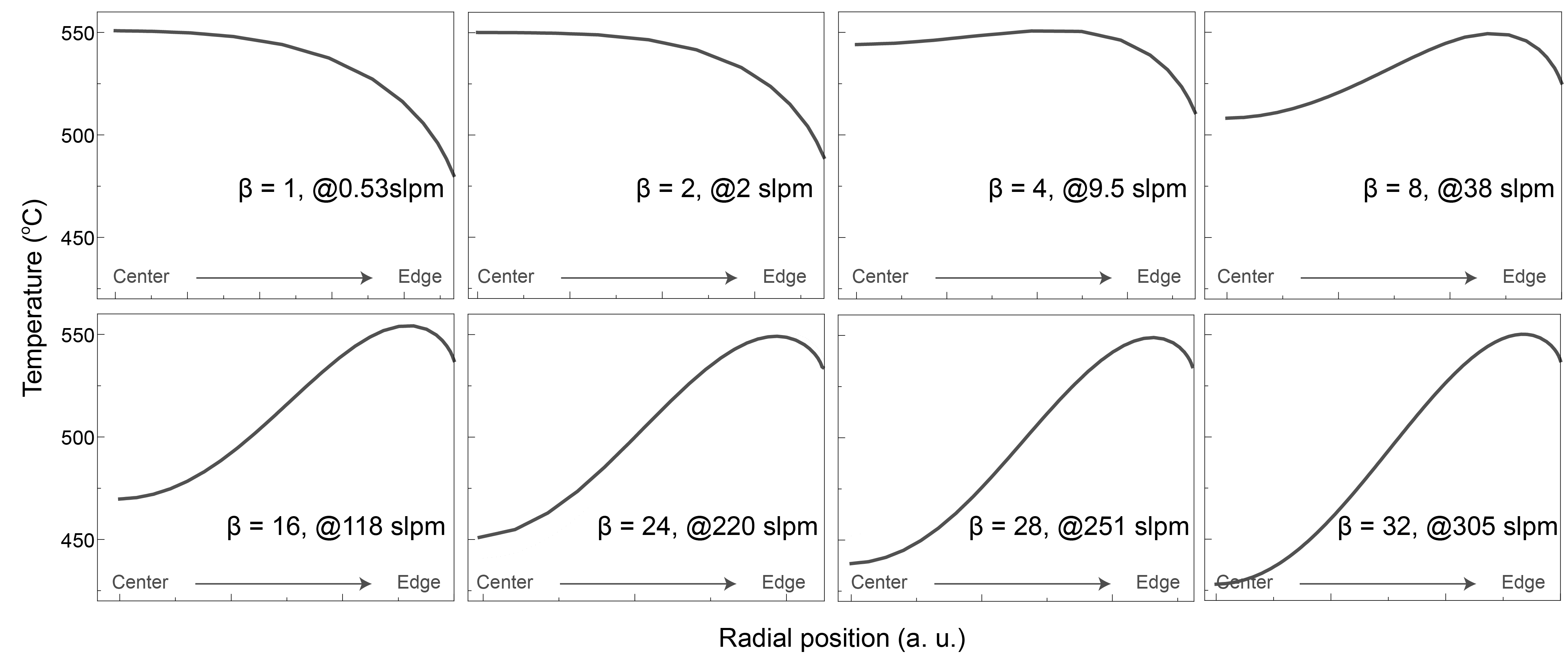}     
  
\textbf{Figure S1.} Simulated radial temperature profiles at the reactor exit for different reactor scaling factors $\beta = 1$ to 32. The inset flow rates are values to achieve 50\% conversion in each case with the exit maximum temperature equal to 550~\,°C.
    \label{fig:fig_s1}
\end{center}

For small reactors ($\beta = 1, 2, 4$), the maximum temperature is near the center with noticeable drops towards the wall due to heat loss through the thermal insulation. However, as reactor size increases, the temperature peak shifts toward the wall, resulting in stronger gradients in the central regime. Although the induction heating profile is volumetric, it is parabolic-like---less heating near the center---leading to cooler cores in large reactors, especially under intensified flow. Obviously, wall-heated reactors become increasingly limited by thermal gradients and are less favorable for scale-up, as shown in Figs. 4 and 5 of the main text.

Note that for $\beta = 4$ in our analysis, the radial temperature profile approaches close to isothermal, allowing us to use the corresponding GHSV as the theoretical limit under a 1D plug-flow assumption. In other words, if such a temperature profile can be preserved at larger scales, the flow rate (in the unit of slpm) increases cubically with $\beta$ and GHSV remains constant, and the maximum efficiency would be eventually set by power loss in the induction system, as shown in Fig.~5B. As shown by the inset values in Fig. S1, for our homogeneous susceptor, though the flow rate keeps increasing as the reactor scales up, it deviates far from the cubical increase rate for larger reactors, resulting in significant drop of the GHSV value. However, ss discussed in the following section, our metamaterial susceptor enables tunable profiles---such as uniform or center-weighted heating---that better support process intensification at scale.

\subsection*{Section S3. Radially tailoring of the heating profile}
\subsection*{S3.1. Guidance by analytic solution: ${1/r^2}$ scaling}
It has been shown that the power density profile within an inductively heated cylinder is a function of its electrical conductivity. Therefore, it is possible to tailor the power density by defining the electrical conductivity as a function of position. It is of unique interest when designing large-scale inductively heated thermochemical reactors to specify a conductivity profile which results in a uniform heating profile. The current density, $\vec{J}$, along the radius of the conductive, nonmagnetic susceptor is, 
\begin{equation}
    \vec{J}(r)=-\frac{kB_0}{\mu_0}\frac{J'_0(kr)}{J_0(kR)}
    \tag{S9}
\end{equation}
Where $k$ is the complex wavenumber, defined as $k=\sqrt{-j\omega\sigma\mu_0}$, $B_0$ is the magnitude of the imposed magnetic field, and for simplicity we assume it as a constant here. $J_0$ is the bessel function of the 0th order, and $R$ is the radius of the cylinder. This description of the current density assumes that the imposed magnetic field has uniform magnitude, and axial and radial shielding effects are negligible. The total power dissipated by the cylinder is, 
\begin{equation}
    P=\frac{1}{2\sigma}\int_V |\vec{J}|^2 dV 
    \tag{S10}
\end{equation}
From this expression it is possible to describe the power density as 
\begin{equation}
p=\frac{1}{2\sigma}|\vec{J}|^2
\tag{S11}
\end{equation}
Substituting the expression for the current density gives a power density of 
\begin{equation}
    p=\Bigg|-\frac{1}{\sqrt{2\sigma}}\frac{kB_0}{\mu_0}\frac{J'_0(kr)}{J_0(kR)}\Bigg|^2
    \tag{S12}
\end{equation}
We will show that specifing a function $\sigma=f(r)$ will produce a constant power density with respect to the radial position, $r$. In particular, it is defined as $\sigma=C/r^2$ where C is a constant. 

Since $k \propto \sqrt{\sigma}$, then with the above definition of $\sigma$, $k\propto \rho$ which means that the argument of $J'_0$ is constant. The power density is simplified to:

\begin{equation}
    p=\Bigg|-\frac{1}{\sqrt{2}}\frac{\sqrt{-j\mu_0}B_0}{\mu_0}\frac{J'_0(\sqrt{-j\mu_0 C})}{J_0(\sqrt{-j\mu_0 C/r^2}R)}\Bigg|^2
    \tag{S13}
\end{equation}
The only term which is a function of $\rho$ is the bessel function of the 0th order in the denominator. A bessel function with an imaginary argument is equal to a modified bessel function with an imaginary prefactor, $J_n(ix)=i^{-n}I_n(x)$. The modified bessel function, $I_n(x)$ goes to 1 as its argument goes to $\infty$. Therefore, the above power density profile will be nearly constant when $C/r^2$ becomes sufficiently large. 

As defined in the main text:
\[
\sigma(r) =
\begin{cases}
\displaystyle \frac{\sigma_{\text{eff}}}{A} \cdot \left( \frac{r}{R} \right)^{-2}, & \text{for } R/5 < r \leq R \\
\\
\displaystyle \frac{\sigma_{\text{eff}}\times 25}{A}, & \text{for } 0 \leq r \leq R/5
\end{cases}
\tag*{(S14)}
\]

Where in this case,
\begin{equation}
    C = \sigma_{\text{eff}}\times R^2/A
    \tag{S15}
\end{equation}

\begin{center}
    \includegraphics[width=0.85\textwidth]{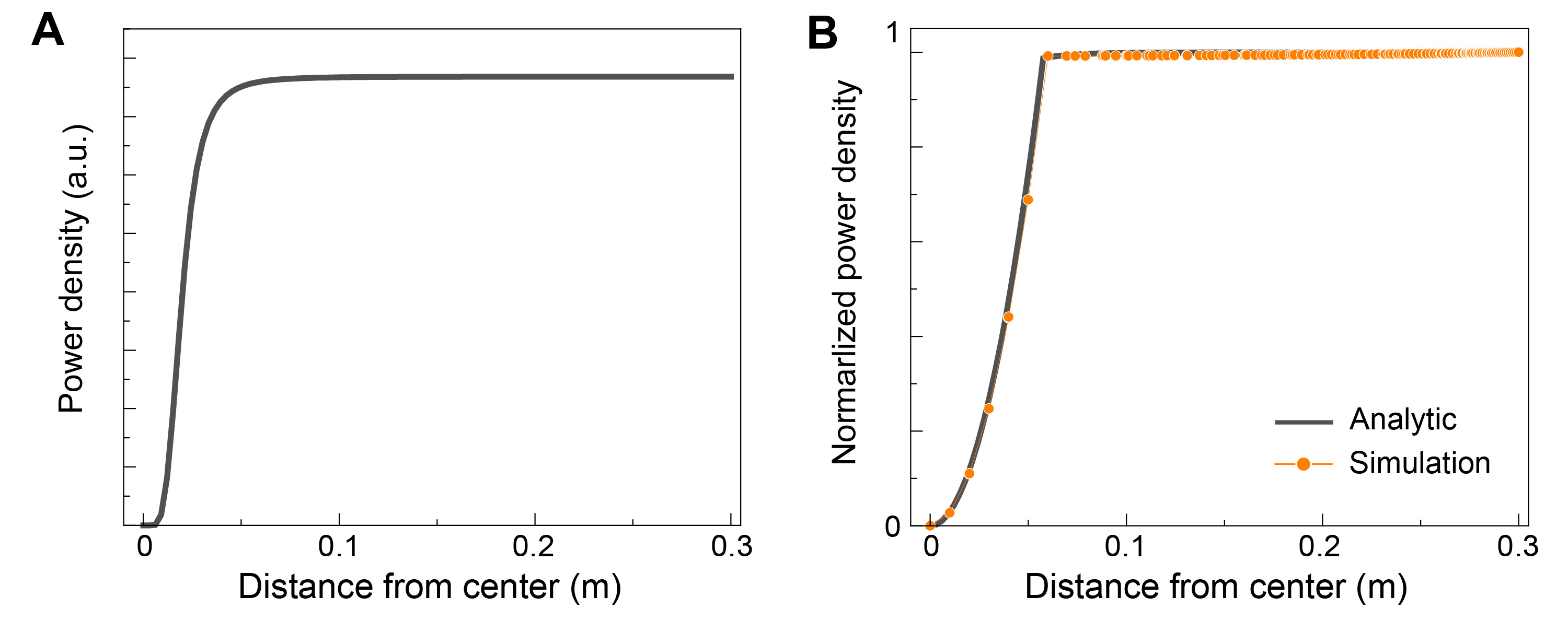} 
    
    \vspace{1ex}
    \justifying
    \textbf{Figure S2.} (A) Power density profile based on analytical expression. (B) Agreement between analytic solution and COMSOL simulation.
    \label{fig:fig_s2}
\end{center}

As shown in Fig.~S2A, for a 0.6-m diameter susceptor, the heating profile can be made nearly uniform by following the analytical solution with constant \( C = 0.001 \, \mathrm{[S \cdot m]} \). This yields near-uniform radial power density across most of the susceptor region. However, power density always drops to zero at the center due to Maxwell boundary conditions. In practice, it is unnecessary to keep increasing conductivity near the center, as such values are beyond material or fabrication limits. Therefore, as defined in the main text, we set the inner region (within radius \( R/5 \)) to a constant conductivity. Although this region contributes less heating, it accounts for only 4\% of the total volume, making it a practical choice. Using this sigma function, we calculated the radial power density via both the analytical solution and COMSOL simulation, which show excellent agreement (Fig.~S2B).

\subsection*{S3.2. Balancing between coupling efficiency and heating uniformity}

It is important to note that while radial tailoring can yield tunable heating profiles that support intensified flows, it may trade off some coupling efficiency. Therefore, identifying the optimal balance between these competing effects for a given reactor configuration and reaction is essential.

In our case, by sweeping the parameter $A$ in Eq.~S14 (also Eq.~1 in the main text), we identified a region where just a few percent sacrifice in coupling efficiency yields a significantly more uniform heating profile. For example, as shown in Fig.~S3A for $\beta = 32$, $A = 7.2$ results in a 93\% coupling efficiency—only 2\% lower than the optimal homogeneous susceptor case—but with much better temperature uniformity. This improves GHSV performance, as shown in Fig.~4 of the main text. Such a sweep is performed across different reactor sizes, with the results summarized in Table~1.

\begin{center}
    \centering
    \includegraphics[width=0.85\textwidth]{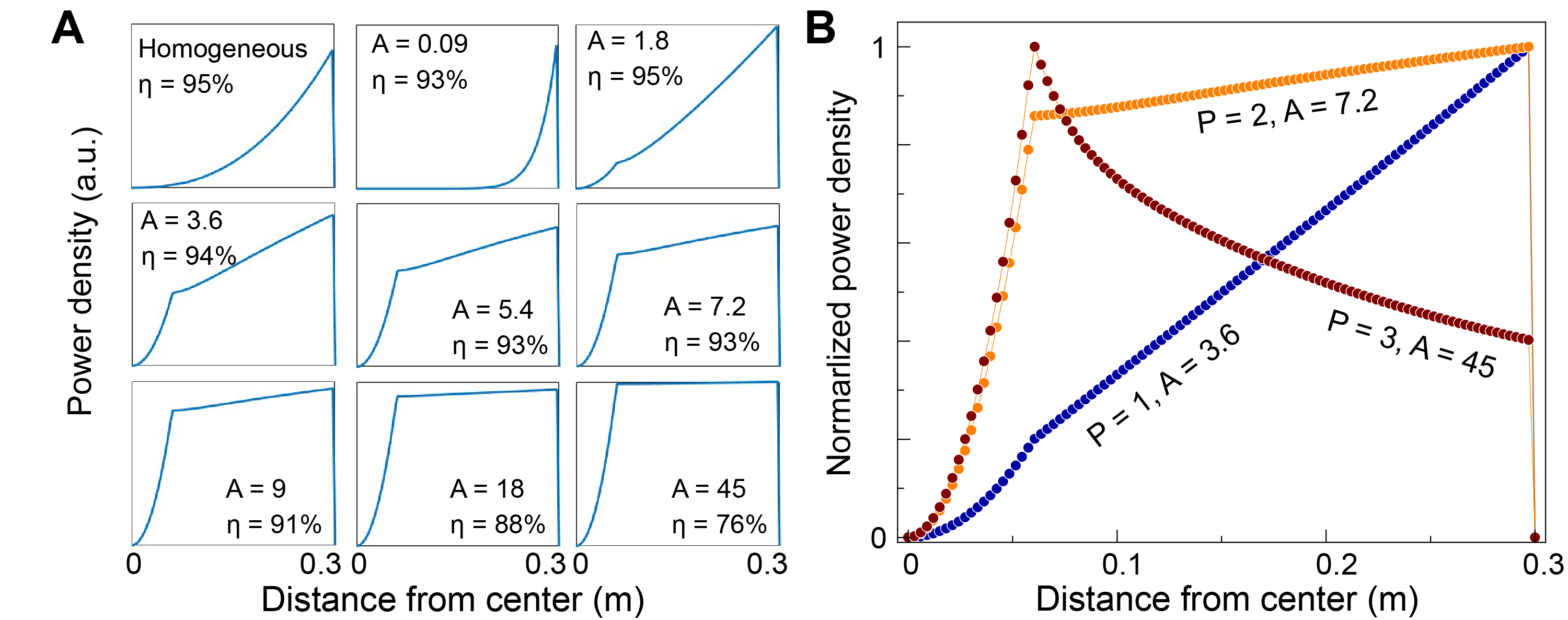}
    
    \vspace{1ex}
    \textbf{Figure S3.} (A) Simulated power density profiles for $\beta = 32$ under various $A$ values in Eq.~1, illustrating the trade-off between heating uniformity and coupling efficiency $\eta$. As $A$ increases, heating becomes more uniform, though at the cost of slightly reduced efficiency. (B) Simulated power density profiles using $1/r^p$ conductivity scaling with different $p$ values, demonstrating tunable heating profiles ranging from center-weighted to edge-weighted. 
    \label{fig:fig_s3}
\end{center}

\subsection*{S3.3. Further tailoring enabled by $1/r^p$ scaling}

Our metamaterial susceptor design enables fully tailorable heating profiles, making it widely applicable to various reactor configurations, thermal properties, and chemical reactions. One additional strategy involves tailoring conductivity using different radial functions of $r$, such as $1/r^p$ with varying $p$ values. This allows realization of diverse heating patterns—such as uniform, center-weighted, or edge-weighted heating—as demonstrated in Fig.~S3B. The results suggest this approach is promising and broadly suitable for inductively heated fixed-bed reactors.

\subsection*{\textbf{Section S4. Estimation of self-resonant frequency of the induction coil}}

To assess and avoid possible electromagnetic interference or energy loss, we estimated the self-resonant frequency (SRF) of the helical induction coils used in our study. 

We implemented the SRF estimation using well-established analytical approximations: Inductance (L) was calculated using Wheeler’s formula for single-layer air-core solenoids:
\begin{equation}
    L \approx \frac{r_c^2 N^2}{9r_c + 10L_c}
    \tag{S16}
\end{equation}

  where $r_c$ is the coil radius, $N$ is the number of turns, and $L_c$ is the coil length. All the dimentional parameters are in the unit of inch.

Capacitance (C) was estimated using Medhurst’s empirical approximation:

\begin{equation}
    C \approx \frac{2\pi D_c}{\cosh^{-1}(p/d_c)}
    \tag{S17}
\end{equation}

  where $D_c$ is the coil diameter, $p$ is the pitch, and $d_c$ is the wire diameter.

The SRF was then determined by:
\begin{equation}
  f_{\text{res}} = \frac{1}{2\pi\sqrt{LC}}
    \tag{S18}
\end{equation}

In practice, to ensure operation well below resonance, one can estimate the cutoff frequency corresponding to a given attenuation (e.g., --30 dB) and quality factor (e.g., $Q = 10$). However, due to the lack of practical $Q$-factor data for scaled reactors, we instead merely identified the SRF regime using the hatched region in Fig.~1B.

\end{document}